%% file: GaugeFrustration.tex
\documentclass[aps, prl, reprint, superscriptaddress]{revtex4-1}

\usepackage{amsmath, amssymb, times, graphicx, mathrsfs, hyperref, array, bbm}
\usepackage[usenames,dvipsnames]{color}

\hypersetup{colorlinks=false}

\graphicspath{{./}{}}

\begin{document}
\title{Thermodynamics of a gauge-frustrated Kitaev spin liquid}

\author{T. Eschmann}
\email[E-mail: ]{eschmann@thp.uni-koeln.de}
\affiliation{Institute for Theoretical Physics, University of Cologne, 50937 Cologne, Germany}
\author{P. A. Mishchenko}
\affiliation{Department of  Applied Physics, The University of Tokyo, Tokyo 113-8656, Japan}
\author{T. A. Bojesen}
\affiliation{Department of  Applied Physics, The University of Tokyo, Tokyo 113-8656, Japan}
\author{Y. Kato}
\affiliation{Department of  Applied Physics, The University of Tokyo, Tokyo 113-8656, Japan}
\author{M. Hermanns}
\affiliation{Department of Physics, Stockholm University, AlbaNova University Center, SE-106 91 Stockholm, Sweden}
\affiliation{Nordita, KTH Royal Institute of Technology and Stockholm University, SE-106 91 Stockholm, Sweden}
\author{Y. Motome}
\affiliation{Department of  Applied Physics, The University of Tokyo, Tokyo 113-8656, Japan}
\author{S. Trebst}
\affiliation{Institute for Theoretical Physics, University of Cologne, 50937 Cologne, Germany}

\begin{abstract}

Two- and three-dimensional Kitaev magnets are prototypical frustrated quantum spin systems, in which the original spin degrees of freedom fractionalize into 
Majorana fermions and a 
$\mathbb{Z}_2$ gauge field -- a purely local phenomenon that reveals itself 
as a thermodynamic crossover at a temperature scale set by the strength of the bond-directional interactions. 
For conventional Kitaev magnets, the low-temperature thermodynamics reveals a second transition at which 
the $\mathbb{Z}_2$ gauge field orders 
and the system enters a spin liquid ground state.
Here we discuss an explicit example that goes beyond this paradigmatic scenario -- the $\mathbb{Z}_2$ gauge field is found to be subject to geometric frustration, the thermal ordering transition is suppressed, and an extensive residual entropy arises. 
Deep in the quantum regime, at temperatures of the order of one per mil of the interaction strength, 
the degeneracy in the gauge sector is lifted by a subtle interplay between the gauge field and the Majorana fermions, 
resulting in the formation of a  Majorana metal.
We discuss the thermodynamic signatures of this physics obtained from large-scale, sign-free quantum Monte Carlo simulations.
\end{abstract}

\maketitle

In frustrated magnetism, lattice gauge theories are an ubiquitous tool to capture the physics of quantum spin liquids \cite{Savary2017quantum}.
The fundamental distinction in these theories between confined and deconfined regimes corresponds to the formation of trivial, magnetically ordered states versus macroscopically entangled spin liquids, respectively.
The principal nature of the underlying gauge theory can further be used to broadly categorize different types of quantum spin liquids such as $\mathbb{Z}_2$ spin liquids \cite{Read1991,Senthil2000}, $U(1)$ spin liquids \cite{AndersonSL}, or chiral spin liquids \cite{KalmeyerLaughlin1987} -- for which the corresponding gauge theory exhibits either a discrete $\mathbb{Z}_2$ or continuous $U(1)$ symmetry or an underlying Chern-Simons action \cite{Wen2002quantum}.
This classification allows one to immediately draw conclusions about the stability of the corresponding spin liquids, in particular
to thermal fluctuations. While the (spontanteous) breaking of time-reversal symmetry in chiral spin liquids mandates their thermal stability and the presence of a finite-temperature phase transition, a more complex picture emerges for $\mathbb{Z}_2$ and $U(1)$ spin liquids. Here, spatial dimensionality needs to be taken into account. For the $\mathbb{Z}_2$ spin liquid
the elementary vison excitations of the underlying gauge structure are point-like objects in two spatial dimensions allowing them to proliferate at finite temperatures and destroy the entangled spin liquid state. In contrast, in three spatial dimensions the  $\mathbb{Z}_2$ spin liquid is stable to thermal fluctuations, as now the visons form (small) loop-like objects that cannot destroy the spin liquid and break open into extended line-like objects only at a finite-temperature transition. 
For $U(1)$ spin liquids the elementary instanton excitations of the bare $U(1)$ gauge theory are point-like objects in both two and three spatial dimensions, implying that these spin liquids are generically not stable at finite temperatures
\footnote{Possible exceptions, guided by physical intuition, are discussed in the context of quantum spin ice \cite{Savary2013}}.

\begin{figure}[b]
    \centering
    \includegraphics[ width=\columnwidth]{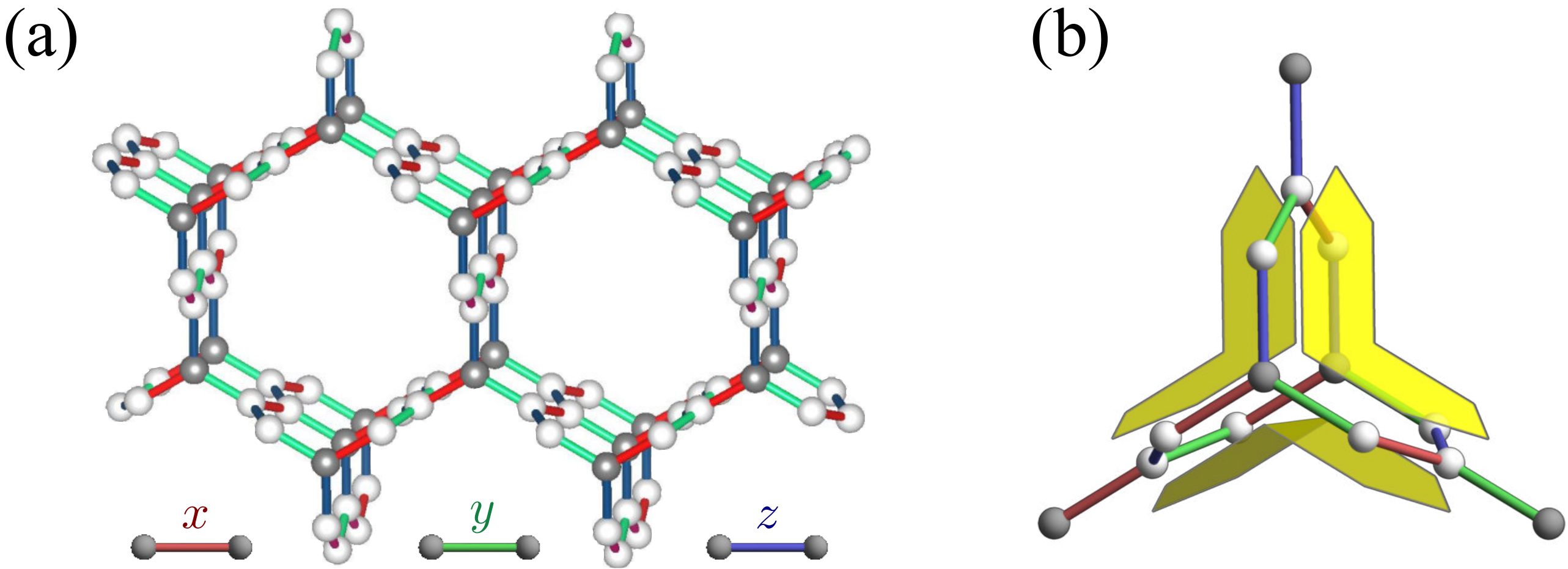}  
      \caption{{\bf The (8,3)c lattice}
      		  illustrated in panel (a) is a three-dimensional tricoordinated lattice, which is built around hexagonal sites (gray shaded)
		  interleaved with six zig-zag chains. Its Schl\"afli symbol (8,3)c indicates that all elementary plaquettes (i.e., loops
		  of shortest length) are of length 8. Around each hexagonal site three such length-8 plaquettes meet as illustrated
		  in panel (b).          
		  }
    \label{fig:Lattice}
\end{figure}

In this manuscript, we consider the explicit example of a three-dimensional $\mathbb{Z}_2$ spin liquid, realized in a 
 numerically tractable Kitaev model, that proves to be an exception from these paradigms. At sufficiently low temperature the gauge field is found to be subject to geometric frustration, arising from local constraints that impose a divergence-free condition on the gauge field 
and ultimately result in an extensive residual entropy. The net result is a suppression of the expected thermal ordering transition of the $\mathbb{Z}_2$ gauge field and the emergence of a spin liquid state that, in some sense, is 
``doubly frustrated", as it arises from the interplay of exchange frustration on the level of the original spin degrees of freedom and geometric frustration on the level of the emerging fractional degrees of freedom. 
Extensive quantum Monte Carlo simulations reveal that
at ultralow temperatures of the order of one per mil of the interaction strength, i.e., deep in the quantum regime, 
the degeneracy in the gauge sector is eventually lifted by a subtle interplay with the Majorana fermion degrees of freedom,
which  emerge in parallel with the gauge field upon fractionalization of the original spin degrees of freedom in Kitaev models. 
The formation of a collective ground state of these itinerant fermions, 
a Majorana metal with a distinct nodal line structure 
feeds back into the gauge sector and leads to the concurrent formation of 
columnar ordering of the gauge field.
Our model system thereby proves to be a principal example of a spin liquid, for which not only the phenomenon of fractionalization, but also of the subsequent non-trivial interplay of the emergent fractional degrees of freedom and the underlying lattice gauge theory can be captured by numerically exact  simulations.

\noindent{\em Gauge frustration.--}      
The Kitaev model with its characteristic bond-directional spin exchanges of the form
\begin{equation}
	\mathcal{H}_{\rm Kitaev} = \sum_{\langle j,k\rangle, \gamma} J_{\gamma} \, \sigma^\gamma_j \sigma^{\gamma}_k \,,
\end{equation}
is well known to be analytically tractable for a class of two-dimensional \cite{Kitaev2006anyons,Hermanns2017physics} and 
three-dimensional \cite{Obrien2016classification} lattice geometries. The analytical treatment relies
on a parton construction introduced by Kitaev that decomposes the spin degrees of freedom into itinerant 
Majorana fermions and a static $\mathbb{Z}_2$ gauge field. 
The fact that the gauge field remains static and assumes, at sufficiently low temperatures, an ordered
ground state is key for the exact solvability of the model, since it allows one to reduce the problem to one 
of free Majorana fermions hopping in a fixed background. 
In fact, a powerful theorem by Lieb \cite{Lieb1994flux} describes the ground state of the gauge sector in terms of $\mathbb{Z}_2$ fluxes through
the elementary plaquettes -- plaquettes of length $6,10, \ldots$ are flux-free, while plaquettes of length 
$4,8, \ldots$ carry a $\pi$ flux. Recent classification work of 3D Kitaev models \cite{Obrien2016classification}
has shown that  Lieb's theorem generically predicts the correct ground state flux assignment, even for lattices that do not fulfill all the mathematical requirements for the theorem to apply.
This is also true for the (8,3)c lattice.

\begin{figure}[b]
    \centering
    \includegraphics[ width=\columnwidth]{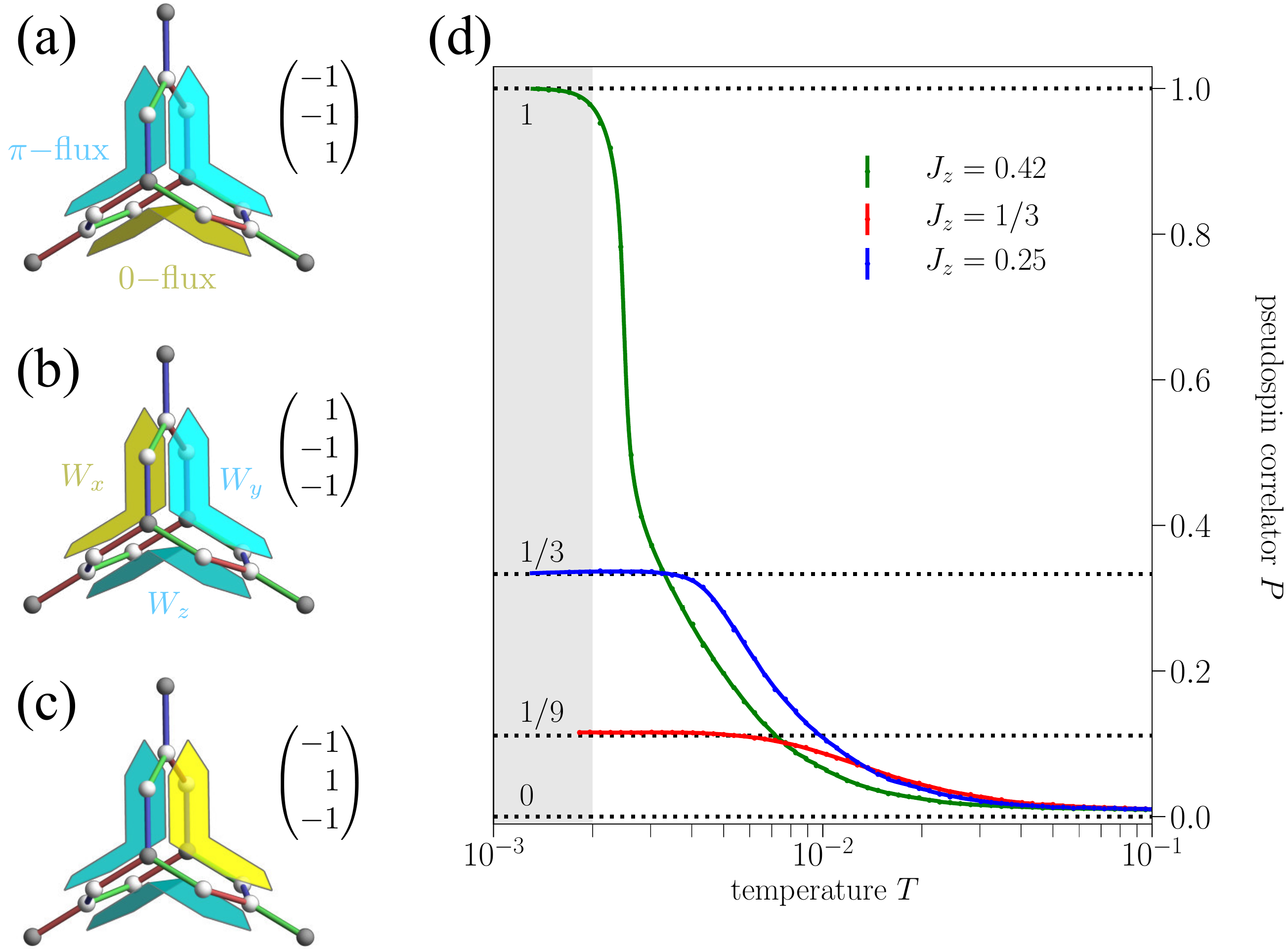}
    \caption{{\bf  Gauge frustration and pseudospins.}
    		 (a)-(c) The three possible 3-flux states that constitute the ground-state manifold of the $\mathbb{Z}_2$ gauge field.
		 The plaquettes are colored cyan/yellow to indicate a $\pi$-/$0$-flux. For each configuration the corresponding 
		 pseudospin vector is specified.
		 (d) Pseudospin correlations as a function of temperature for different coupling strengths.
		Data shown is for system size $4\cdot4\cdot6$.			
		The gray shaded area indicates the temperature region in which we have employed histogram reweighting techniques
			\cite{FerrenbergSwendsen88,FerrenbergSwendsen89}
			to extrapolate the data.
          }
    \label{fig:PseudoSpins}
\end{figure}

The key ingredient for the study at hand is a ``frustrated" three-dimensional lattice geometry whose
central motif are hexagonal sites at which three plaquettes of length 8 meet, see the illustration in Fig.~\ref{fig:Lattice}. 
Following the above intuition based on Lieb's theorem each of these plaquettes is destined to carry a $\pi$  flux, 
which however conflicts with the fact that for any closed volume, such as the one spanned by the three neighboring plaquettes,
the fluxes must obey a divergence-free condition -- if a flux enters the volume through one of the
plaquettes, it must leave through another one. This divergence-free condition allows only two of the three
plaquettes to carry a $\pi$
 flux and leaves one of the plaquettes in a flux-free state. For isotropic 
coupling strength $J_x = J_y = J_z$  this produces three possible flux arrangements per hexagonal site and 
an extensive residual entropy for the entire system. It is this formation of an extensive manifold of (almost) degenerate states
in the gauge sector that designates the term ``gauge-frustrated" for the Kitaev model at hand
\footnote{{A similar scenario of ``gauge frustration" plays out in the 3D Kitaev model on the hypernonagon lattice \cite{Kato2017}
	where the pattern of $\pm \pi/2$ fluxes arising from elementary 9-bond loops is subject to geometric frustration.}}.

One way to relieve the frustration in the system is to vary the relative coupling strengths of the bond-directional 
exchange (keeping an overall normalization $J_x+J_y+J_z=1$). To see this, consider that every length-8 plaquette consists of an uneven number of bond-directional coupling types, 
e.g., $3 \times J_x, 3 \times J_y, 2 \times J_z$ for the bottom plaquette illustrated in Figs.~\ref{fig:PseudoSpins}(a)-(c),
while the two upper plaquettes have $3 \times J_z$ couplings.
If the $z$-bond coupling is enhanced, i.e. for $J_z > J_x = J_y$, one finds that the local threefold degeneracy is immediately lifted
and only one local gauge configuration, illustrated in Fig.~\ref{fig:PseudoSpins}(a) is favored. 
For $J_z < J_x = J_y$  the two flux configurations of Figs.~\ref{fig:PseudoSpins}(b),(c) remain degenerate, thus only 
partially lifting the original threefold degeneracy.

To check that this phenomenon of gauge frustration, 
which in the above line of arguments is primarily motivated by the intuition build on Lieb's theorem, 
indeed plays out in the model at hand, 
we have performed large-scale sign-free Monte Carlo simulations of the model over a wide range of temperatures. 
To capture the local gauge physics, we define for any given hexagonal site a pseudospin vector 
\begin{equation}
   {\bf W} = 
  \begin{pmatrix} W_x \\ W_y \\ W_z \end{pmatrix}
  		     \stackrel{\rm (a)}{=}  \begin{pmatrix}  - 1 \\ -1 \\ \phantom{-}1 \end{pmatrix}
  		     \stackrel{\rm (b)}{=}  \begin{pmatrix} \phantom{-}1 \\ - 1 \\ -1 \end{pmatrix}
  		     \stackrel{\rm (c)}{=}  \begin{pmatrix}  - 1 \\ \phantom{-}1 \\ -1 \end{pmatrix}  \,,
\end{equation}
where the individual components $W_{x,y,z}$ can take values +/- 1 indicating the absence/presence of 
a $\pi$ 
flux in the three adjacent plaquettes, thus allowing for eight different possible vectors. 
For the three states that fulfill the local divergence-free condition, their $\pi$ flux 
assignments are given on the r.h.s.~of the above equation in correspondence
with Figs.~\ref{fig:PseudoSpins}(a)-(c).
Using these pseudospin vectors we can define a two-point correlation function
\begin{equation}
	P = \frac{4}{3N} \sum_j \langle {\bf W}_0 \cdot {\bf W}_j\rangle  \,,
	\label{eq:PseudospinCorrelations}
\end{equation}
where $0$ and $j$ denote two hexagonal sites of the lattice.
$P$ readily reveals the nature of the ground-state manifold 
and can be directly probed in our Monte Carlo simulations.
Its expectation value is 
$P = 1$ for the case of a single ground-state of the gauge field ($J_z > J_x, J_y$),
and $P<1$ for the extensively degenerate cases, specifically
$P = 1/3$ for the scenario with a local twofold degeneracy ($J_z < J_x, J_y$) and 
$P = 1/9$ for the scenario with a local threefold degeneracy expected for isotropic coupling strengths ($J_z = J_x = J_y$)
-- see the Supplemental Material for an analytical derivation.
Numerical results from Monte Carlo runs are shown in Fig.~\ref{fig:PseudoSpins}(d) for different strengths of $J_z$.
The data clearly shows that down to temperatures of the order of $10^{-2}$ (in units of the Kitaev coupling $J$) 
the pseudospin correlation function goes to zero, indicating a completely disordered state of the gauge fields. Below
this temperature scale, the pseudospin correlation function rises and indeed saturates.
These simulations thereby unambiguously confirm that the system indeed enters a regime of gauge frustration at low
temperatures, with an extensive degeneracy building up in the gauge sector.

\begin{figure}[b]
    \centering
    \includegraphics[ width=\columnwidth]{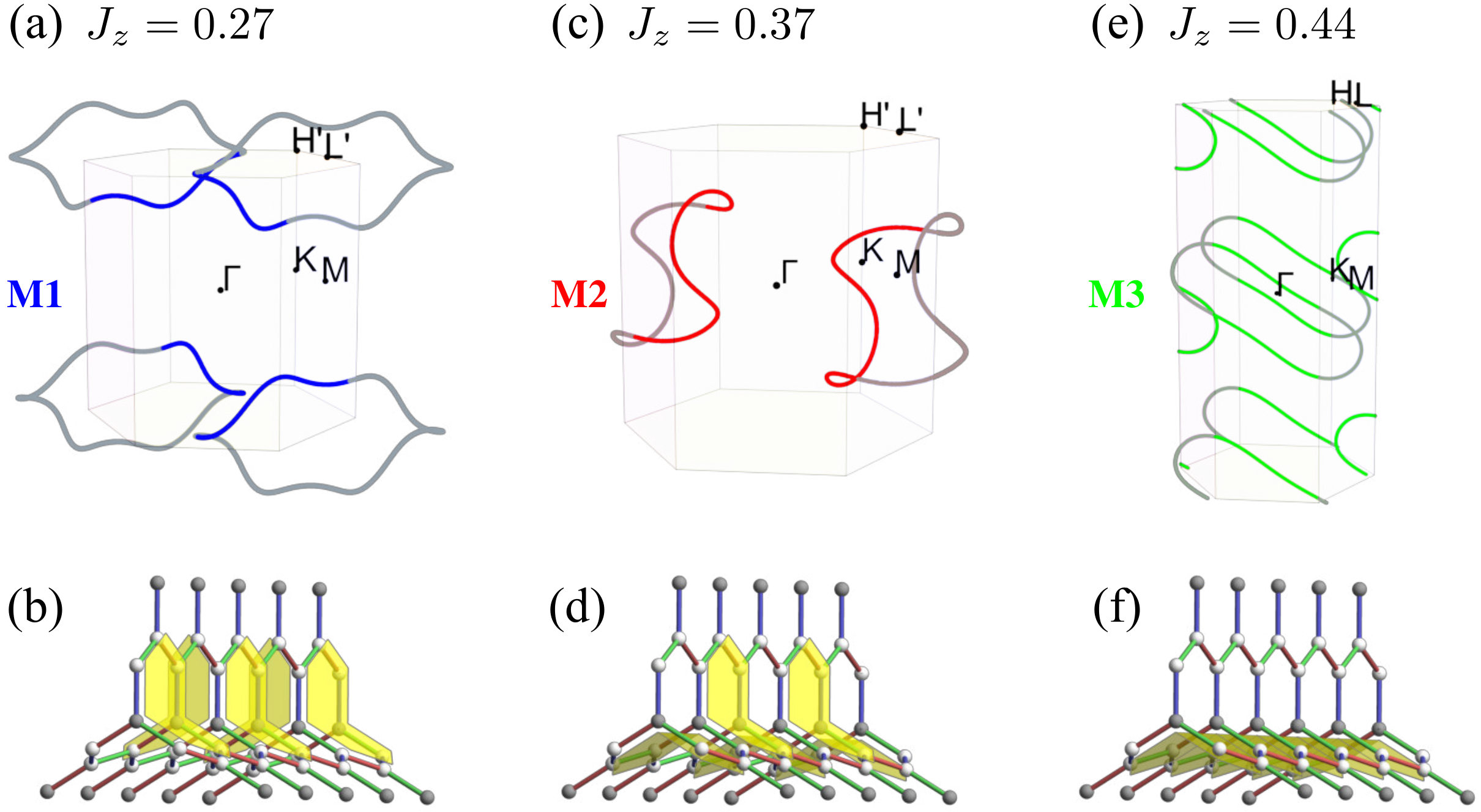}\\
        \caption{{\bf Majorana semimetals and columnar gauge ordering.} 
        			Evolution of the nodal line in the Majorana band structure for varying coupling $J_z$ (top row)
			calculated for the columnar-ordered gauge field configurations illustrated in the bottom row.
          }
    \label{fig:nodal}
\end{figure}

\paragraph{Lifting of gauge degeneracy.--}

The formation of an ``accidential" degeneracy, 
i.e., a degeneracy that is not protected by any inherent symmetries of the system, 
 is often accompanied by some residual effect that splits this degeneracy, at sufficiently small temperature scales, 
 in favor of a unique (or less degenerate) ground state 
 -- an  effect that typically goes hand-in-hand with a macroscopic phase transition. 
Such residual effects can include the energetic or entropic selection of ground states, driven either by 
otherwise negligible interactions (such as e.g., longer-range interactions) or thermal fluctuations in 
an order-by-disorder scheme \cite{Villain1980}.

For the Kitaev system at hand, we find the particularly intriguing scenario that it is an (energetic) interplay between
the emergent fractional degrees of freedom that ultimately lifts the gauge frustration discussed above.
From the perspective of the itinerant Majorana fermions, the residual degeneracy in the gauge sector is equivalent to
a complex scattering potential, as every individual gauge configuration corresponds to a distinct sign structure of the 
Majorana hopping amplitudes. In the gauge frustrated regime, the collective state of the Majorana fermions is therefore 
best described as a thermal metal \cite{Chalker1988,Chalker2001}, as the degeneracy in the gauge sector has a similar effect as (thermal) disorder.
This observation readily points to a scenario where the formation of a collective Majorana state -- a more conventional, 
disorder-free metallic state -- might become favorable at the expense of inducing an ordering in the gauge sector. 
This is precisely what happens at ultralow temperatures, of the order of $10^{-3}$ of the magnetic coupling strength, 
in the system at hand -- the itinerant Majorana degrees of freedom form a  nodal-line semimetal, while simultaneously enforcing 
a columnar ordering in the gauge sector that lifts the gauge frustration. Schematically, the key signatures of these states are 
illustrated in Fig.~\ref{fig:nodal}, which shows the gapless nodal line in the Majorana band structure for different values of
the bond-directional exchange $J_z$, and the corresponding columnar ordering patterns of the gauge field.

\begin{figure}[t]
    \centering
    \includegraphics[ width=.95\columnwidth]{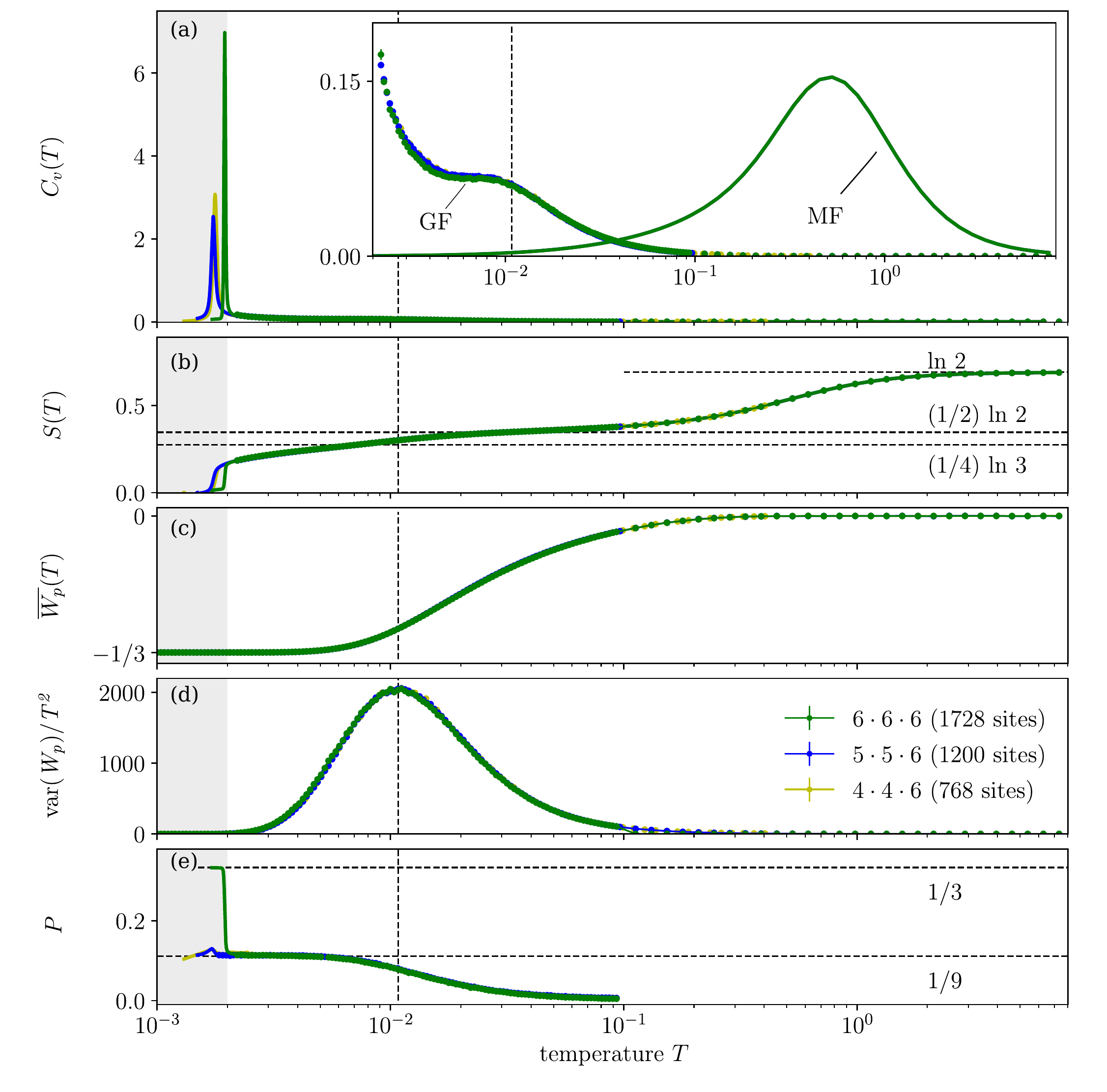}\\
        \caption{{\bf Thermodynamic signatures} for the isotropic system. 
        			(a) Specific heat, separated into  contributions of  $\mathbb{Z}_2$ gauge field (GF) 
				and itinerant Majorana fermions (MF). 
			(b) Entropy per spin.
			(c) Flux per plaquette $\overline{W_p}= \sum_p W_p/N$.
			(d) Fluctuation of the flux per plaquette $(\langle W_p^2 \rangle - \langle W_p \rangle^2) / T^2$.
			(e) Pseudospin correlator \eqref{eq:PseudospinCorrelations}. 
			The dashed line indicates the temperature scale at which the system enters the constrained manifold, 
			corresponding to the maximum in the flux fluctuations and a residual entropy of $(1/4) \ln 3$. 
			Error bars are smaller than the symbol sizes.
			The gray shaded area indicates the temperature region in which we have employed histogram reweighting techniques
			\cite{FerrenbergSwendsen88,FerrenbergSwendsen89} to extrapolate the data.
          }
    \label{fig:thermo}
\end{figure}

\paragraph{Thermodynamics.--}      

To quantitavily probe this physics we have measured a variety of thermodynamic observables in quantum Monte Carlo (QMC) simulations covering four orders of magnitude in temperature.
These QMC simulations are performed in the sign-free parton basis \cite{Nasu2014vaporization}, 
i.e.~we sample configurations of the gauge field, $\{u_{jk} = \pm 1\}$ for every bond $\langle j,k \rangle$ of the lattice,  
with the change of the Majorana free energy, $F_f(\{u_{jk} \})$, being calculated
explicitly in every update step (either by exact diagonalization or a Green's function based kernel polynomial method \cite{Weisse2009,Mishchenko2017}, see the Supplemental Material for further details of this implementation).
 This procedure also allows us to separately distill the entropic contributions to the specific heat of the Majorana fermions \cite{shimomura_spin_ice}
\begin{equation}
C_{\it v, \rm MF}(T) = -\frac{1}{T^2}  \left \langle \frac{\partial E_f(\{u_{jk}\})}{\partial \beta} \right \rangle_{\rm MC}  \,,
\end{equation}
and the gauge field
 \begin{equation}
C_{\it v, \rm GF}(T) = \frac{1}{T^2} \left(\left \langle E_f^2(\{u_{jk} \}) \right \rangle_{\rm MC} - \left \langle E_f(\{u_{jk} \}) \right \rangle_{\rm MC}^2 \right ) \,.
\end{equation}
Results are given in Fig.~\ref{fig:thermo}(a) for the isotropic coupling point ($J_x=J_y=J_z$). Some features
of the multipeak structure of the specific heat are well known from conventional Kitaev models, such as the crossover feature
at temperatures of order 1, where the system releases about half of its entropy, see Fig.~\ref{fig:thermo}(b), upon the fractionalization of the local spin degrees of freedom \cite{Nasu2014vaporization,Nasu2015thermal}, primarily by the Majorana fermions (whose energy scale is set by the hopping/magnetic coupling strength). 
Below this crossover peak there are two additional features in the specific heat that originate in the gauge sector. 
At a temperature of about $10^{-2}$ a broad shoulder forms, which does not show any scaling with system size pointing to 
a local crossover phenomenon. It is at this temperature scale that the system enters the manifold of ``gauge frustrated" states,
which is evident from (i) the average plaquette flux dropping to a value of $\langle W_p \rangle = -1/3 = (-2 +1)/3$ 
(expected for local configurations where two out of three plaquettes have a $\pi$ 
flux, i.e., $W_p=-1$, and one plaquette
remains flux-free, $W_p=+1$), (ii) the fluctuations of the plaquette flux exhibiting a maximum upon entering this constrained ground-state manifold, 
and (iii) the pseudospin correlator \eqref{eq:PseudospinCorrelations} raising and saturating at the expected value of $P=1/9$, 
as documented in Figs.~\ref{fig:thermo}(c)-(e).
Below this second crossover peak, at a temperature of the order of $2\times 10^{-3}$, one finds a sharp peak in the specific heat
that sharpens with increasing system size -- this is a true thermal phase transition, where the system releases entropy by forming
a columnar ordering of the gauge field. In this ordered state every column of hexagonal sites exhibits a staggered pattern of
the flux-free plaquettes as indicated by the yellow plaquettes in Fig.~\ref{fig:nodal}(b). However, since the columns order individually
and there are two possible staggered states for each column, the resulting overall order not only still allows for a residual entropy, but
also breaks the lattice rotational symmetry. This is a remarkable symmetry-breaking effect as it plays out {\em solely} in the gauge sector.

\begin{figure}[t]
   \centering
    \includegraphics[width=\columnwidth]{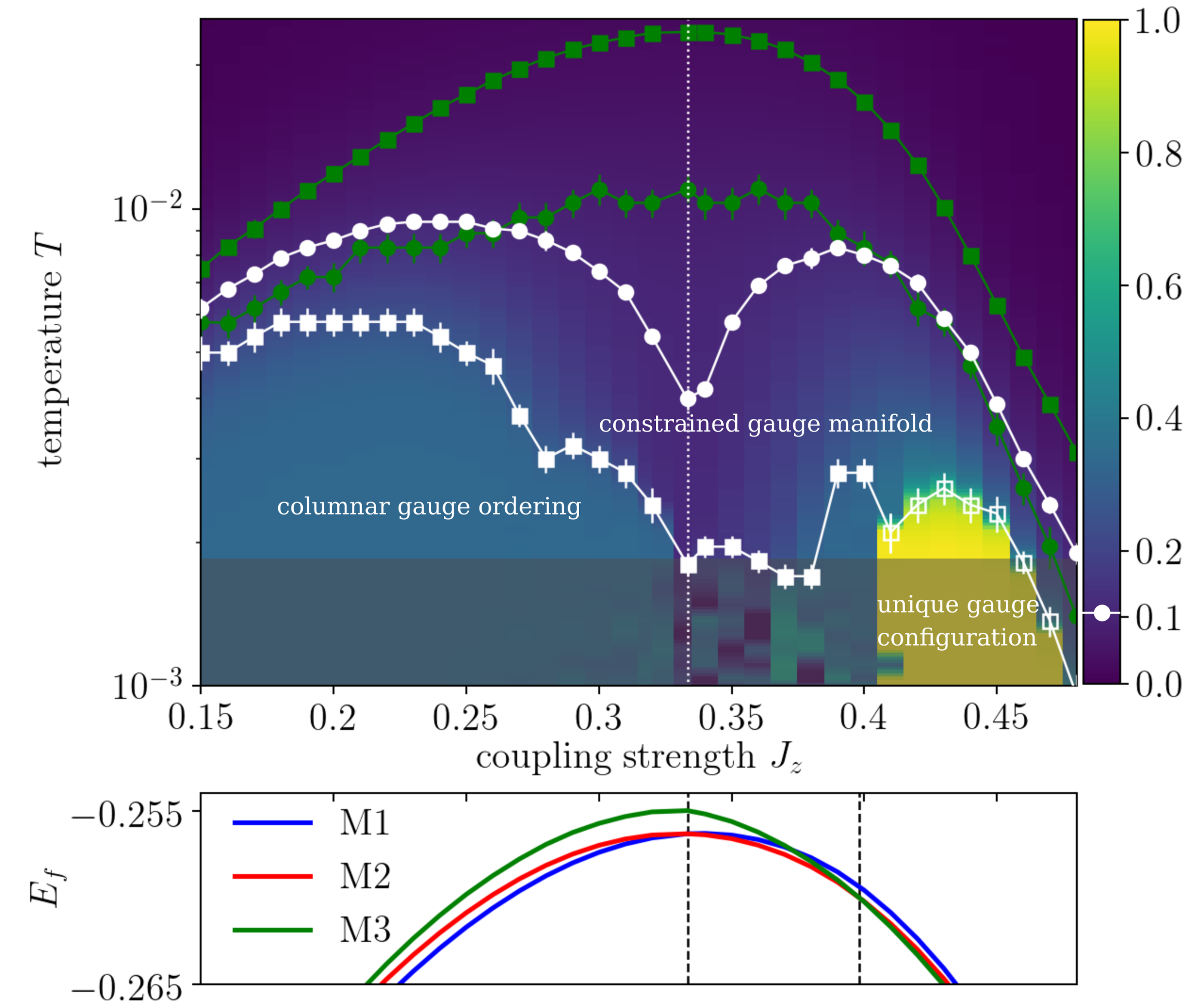}  
   \caption{
     {\bf Finite-temperature phase diagram.} 
	The color-coding of the background contour plot indicates the pseudospin correlations \eqref{eq:PseudospinCorrelations}
	as a function of temperature $T$ and coupling strength $J_z$. 
	The crossover scale at which the system enters the  flux constrained manifold of the gauge field
	is indicated by (i) the filled green circles indicating the peak in the variance of the fluxes
	and (ii) filled white circles indicating temperature points for which the pseudospin correlator $P = 1/9$.
	The onset of flux ordering is signaled in the high-temperature regime where we mark the line along which the flux
	becomes $\langle W_p \rangle = - 1/6$ (green squares). The low-temperature columnar ordering transition of the gauge field is
	marked by the white squares (indicating the location of the corresponding diverging peak in the specific heat).
	Filled squares indicate transitions where the degenerate manifold of constrained gauge configurations is 
	lifted by the formation of a Majorana metal, while open squares indicate transitions driven by an energetic
	selection within the gauge sector. Data shown is for system size $4\cdot4\cdot6$.
	The  shaded area indicates the temperature region in which we have employed histogram reweighting techniques
	\cite{FerrenbergSwendsen88,FerrenbergSwendsen89} to extrapolate  data from parallel tempering simulations \cite{Hukushima1996}.
	The lower panel shows the ground-state energy of the nodal-line semimetals for the 3 types of columnar gauge ordering
	discussed in the main text, using the same color code as in Fig.~\ref{fig:nodal}.
   }
    \label{fig:qmc}
\end{figure}

\paragraph{Phase diagram.--}      

Expanding this analysis of key thermodynamic observables to a range of $J_z$ parameters, we have compiled
the composite phase diagram of Fig.~\ref{fig:qmc}. Plotted here are different indicators for the crossover scale 
to the constrained gauge manifold.
As proxies for this crossover, we have marked (i) the location of the peak in the variance of the fluxes, akin to Fig.~\ref{fig:thermo}(d), 
by the filled green circles
and (ii) the line of temperature points at which the pseudospin correlator \eqref{eq:PseudospinCorrelations} crosses $P=1/9$
by the filled white circles. 
The low-temperature phase transition, at which the concurrent formation of a nodal-line Majorana semimetal 
and columnar order of the gauge field occurs, is indicated by the white squares. Depending on the strength of $J_z$, 
we distinguish two principal scenarios. First, there is a line of transitions (indicated by the filled white squares) where it is 
the formation of the Majorana metal that lifts the degeneracy in the gauge sector and enforces the columnar gauge order.
This is the case for $J_z \leq 1/3$. For $J_z \gtrsim 0.40$, it is the energetics within the gauge sector that readily selects
a single configuration of the constrained gauge field for each hexagonal site, see Fig.~\ref{fig:PseudoSpins}(a), which results in the
columnar ordering depicted in Fig.~\ref{fig:nodal}(f). For $1/3 < J_z \lesssim 0.40$, a more subtle mechanism is at play
where the energetics of the gauge field favors the same type of columnar order as for $J_z \gtrsim 0.40$, but the minimization
of the Majorana energy enforces yet another type of columnar order, depicted in Fig.~\ref{fig:nodal}(d), which, in a certain sense,
is an intermediate type of order with a staggered flux pattern involving flux-free states on some of the bottom $W_z$ plaquettes.
The overall phase diagram thereby reveals multiple distinct regimes, in which a subtle interplay between the emergent
parton degrees of freedom leads to the formation of different types of collective ground states -- including gapless spin liquids with
a Majorana nodal line and columnar-ordered $\mathbb{Z}_2$ gauge fields.

\paragraph{Conclusions.--}      

The main results of the study at hand are two advances in the conceptual understanding of quantum spin liquids. First, we have introduced the concept of ``gauge frustration", which we showcased in a three-dimensional Kitaev model where the emergent $\mathbb{Z}_2$ gauge degrees of freedom are subject to local constraints resulting in an extensive residual entropy. Second, we showed by large-scale numerical simulations that this residual entropy can be lifted by an interplay of the $\mathbb{Z}_2$ lattice gauge theory and the itinerant Majorana fermions, which concurrently emerge with the gauge field upon fractionalization of the original local spin degrees of freedom. As such, the model at hand realizes a  scenario intermediate between more conventional Kitaev models where the parton degrees of freedom fully decouple (allowing for an analytical solution where one first identifies the ground state of the gauge field and subsequently solves the Majorana problem), and the scenario of strongly interacting partons as it is the case for, e.g., a $U(1)$ spin liquid, in which the gauge field remains heavily fluctuating to the lowest temperatures and thereby strongly feeds back into the formation of a collective parton state.

\acknowledgments
{\em Acknowledgments.--}
T.E., M.H., and S.T. acknowledge partial funding by the Deutsche Forschungsgemeinschaft (DFG, German Research Foundation) 
-- Projektnummer 277146847 -- SFB 1238 (projects C02 and C03).
M.H. acknowledges partial funding by the Knut and Alice Wallenberg Foundation and the Swedish Research Council. 
P.A.M, T.A.B, Y.K., and Y.M. acknowledge funding by Grant-in-Aid for Scientific Research under Grant No. 15K13533 and 16H02206. Y.M. and Y.K. were also supported by JST CREST (JPMJCR18T2).
The numerical simulations were performed on the JUWELS cluster at the Forschungszentrum J\"ulich.

\vskip -2mm

\bibliography{./Kitaev.bib}

\newpage
\input{supp.tex}

\end{document}

%% file: supp.tex
\widetext

\begin{center}
  \textbf{\large Supplemental Material}
\end{center}

\setcounter{section}{0}
\setcounter{equation}{0}
\setcounter{figure}{0}
\setcounter{table}{0}
\setcounter{page}{1}
\makeatletter
\renewcommand{\theequation}{S\arabic{equation}}
\renewcommand{\thefigure}{S\arabic{figure}}

\section{Lattice structure}
The lattice vectors of the inversion-symmetric (8,3)c lattice are chosen as
\begin{align*}
\textbf{a}_1 = \left(1, 0, 0 \right),~~~~
\textbf{a}_2 = \left(-\frac{1}{2}, \frac{\sqrt{3}}{2}, 0 \right),~~~~
\textbf{a}_3 = \left(0, 0, \frac{2}{5} \right)
\end{align*}
and the reciprocal lattice vectors become
\begin{align*}
\textbf{b}_1 = 2\pi\left(1, \frac{1}{\sqrt{3}}, 0 \right),~~~~
\textbf{b}_2 = 2\pi\left(0, \frac{2}{\sqrt{3}}, 0 \right),~~~~
\textbf{b}_3 = 2\pi\left(0, 0, \frac{5}{2} \right).
\end{align*}
It has eight sites per unit cell that are located at
\begin{align*}
\textbf{r}_1 &= \left(-\frac{1}{5}, \frac{4}{5\sqrt{3}}, \frac{1}{10} \right),~~~~&&
\textbf{r}_2 = \left(0, \frac{7}{5\sqrt{3}}, \frac{1}{10} \right),~~~~&&
\textbf{r}_3 = \left(\frac{1}{5}, \frac{4}{5\sqrt{3}}, \frac{1}{10} \right),~~~~&&
\textbf{r}_4 = \left(\frac{1}{2}, \frac{1}{2\sqrt{3}}, \frac{3}{10} \right), \\
\textbf{r}_5 &= \left(0, \frac{1}{\sqrt{3}}, \frac{1}{10} \right),~~~~&&
\textbf{r}_6 = \left(\frac{3}{10}, \frac{7}{10\sqrt{3}}, \frac{3}{10} \right),~~~~&&
\textbf{r}_7 = \left(\frac{1}{2}, \frac{1}{10\sqrt{3}}, \frac{3}{10} \right), ~~~~&&
\textbf{r}_8 = \left(\frac{7}{10}, \frac{7}{10\sqrt{3}}, \frac{3}{10} \right).
\end{align*}

\begin{figure}[h]
    \centering
    \includegraphics[ width=.5\columnwidth]{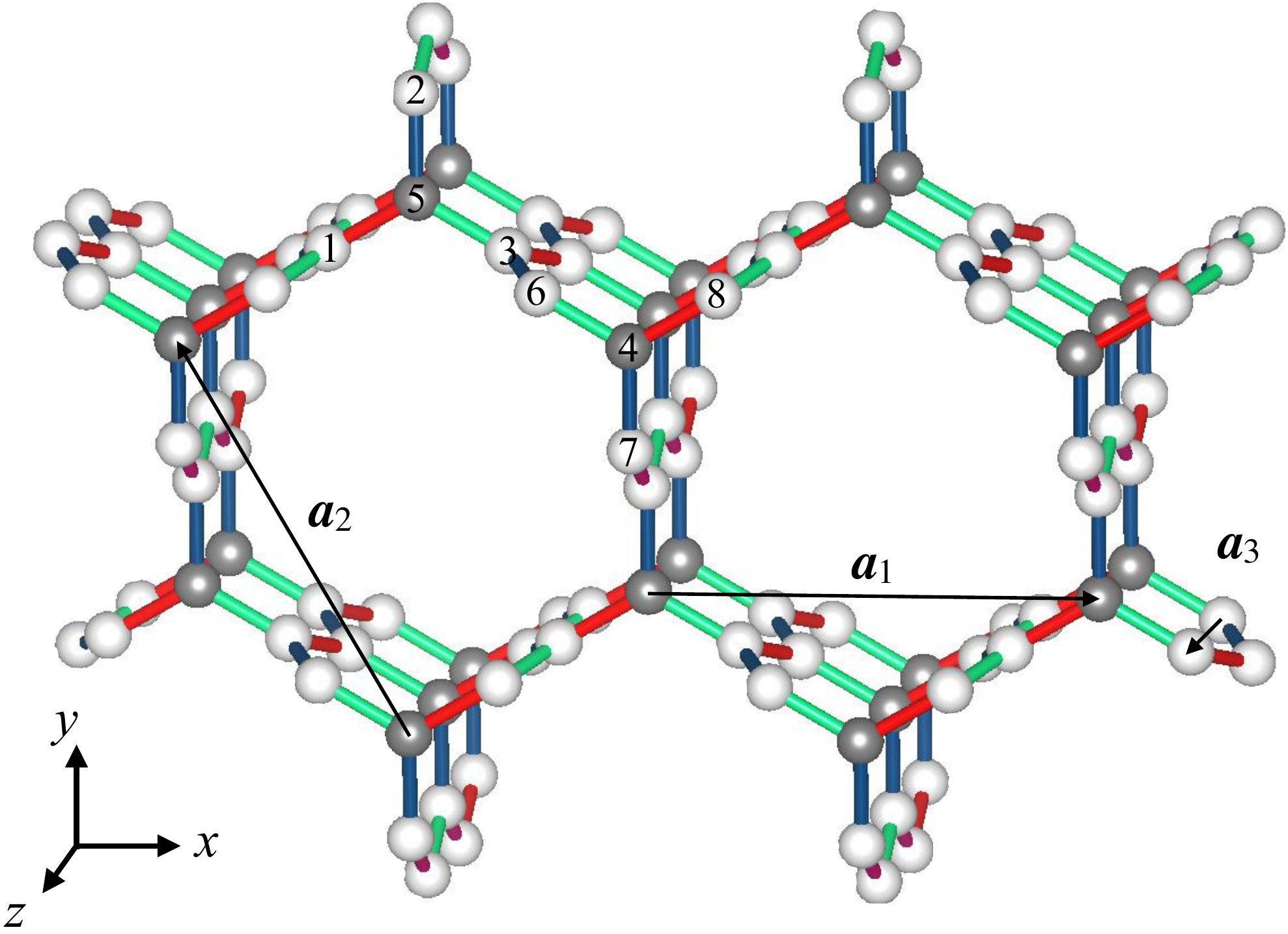}  
      \caption{{\bf The unit cell of the (8,3)c lattice} consists of eight sites as indicated in the figure. 
      			Also indicated are the three lattice vectors.     
		  }
    \label{fig:UnitCell}
\end{figure}

\section{Pseudospin correlations}

The pseudospin correlations defined in \eqref{eq:PseudospinCorrelations} can be used to identify the ground state manifolds of the different (an)isotropic regimes of the system. 

We can introduce the unitary transformation \cite{Ishizuka2013,Fujiki1984} 
\begin{equation}
{\bf P} = \left( \begin{array}{ccc}
\frac{2}{\sqrt{6}} & -\frac{1}{\sqrt{6}} & -\frac{1}{\sqrt{6}} \\ 
0 & \frac{1}{\sqrt{2}} & -\frac{1}{\sqrt{2}} \\ 
\frac{1}{\sqrt{3}} & \frac{1}{\sqrt{3}} & \frac{1}{\sqrt{3}}
\end{array} \right) \left( \begin{array}{c}
W_x \\ 
W_y \\ 
W_z
\end{array} \right)\,,
\label{eq:PseudospinDefinition}
\end{equation}
mapping the triplet of 3-flux states that preserve the local divergence-free condition to be rotationally symmetric around the $W_z$-axis: 

\begin{equation}
{\bf P}_a =  \left( \begin{array}{c} -\sqrt{\frac{2}{3}} \\ -\sqrt{2} \\ -\frac{1}{\sqrt{3}} \end{array} \right), \hspace{0.2cm}
{\bf P}_b =  \left( \begin{array}{c} 2 \sqrt{\frac{2}{3}} \\ 0 \\ -\frac{1}{\sqrt{3}} \end{array} \right), \hspace{0.2cm}
{\bf P}_c =  \left( \begin{array}{c} -\sqrt{\frac{2}{3}} \\ \sqrt{2} \\ -\frac{1}{\sqrt{3}} \end{array} \right) \,.
\label{eq:GS_pseudospins}
\end{equation}

It can be easily seen that $\|{\bf W}_m\| = \|{\bf P}_m\| = \sqrt{3}$ for all $m \in \{a,b,c\}$ and

\begin{equation}
\frac{{\bf P}_m \cdot {\bf P}_n}{\|{\bf P}_m\|  \|{\bf P}_n\|} = \frac{{\bf W}_m \cdot {\bf W}_n}{\|{\bf W}_m\|  \|{\bf W}_n\|} = \left\{ \begin{array}{c}-\frac{1}{3}, \hspace{0.2cm} m \neq n \\ 1, \hspace{0.5cm} m = n \end{array} \right. 
\end{equation}

For large $T$, all eight possible 3-flux states are allowed, so we expect P to average to 0. For the low temperature regime of the isotropic system ($J_x = J_y = J_z$), the pseudospins ${\bf P}_a$, ${\bf P}_b$, ${\bf P}_c$ will be equally distributed in the system, giving the expectation value of the pseudospin correlations
\begin{equation}
P_{abc} = \frac{1}{9} \left(1 + 1 + 1 - 6 \cdot \frac{1}{3}\right) = \frac{1}{9} \,.
\label{eq:PseudospinExample1}
\end{equation}
For the weak-$J_z$ limit, only the pseudospins ${\bf P}_b$, ${\bf P}_c$ are selected, resulting in
\begin{equation}
P_{bc} = \frac{1}{4} \left(1 + 1 - 2 \cdot \frac{1}{3}\right) = \frac{1}{3} \,,
\label{eq:PseudospinExample2}
\end{equation}
while for the strong-$J_z$ limit, all plaquette triplets select the state ${\bf P}_a$, leading to
\begin{equation}
P_{a} = 1 \,.
\label{eq:PseudospinExample3}
\end{equation}

Since we choose all our systems to have periodic boundary conditions in all spatial directions, each lattice with $N$ sites has $N_p = 3N/4$ plaquettes and $N_s = N/4$ pseudospins. For a finite system, the pseudospin correlator results as

\begin{equation}
	P = \frac{1}{N_s} \sum_j \frac{\langle {\bf W}_0 \cdot {\bf W}_j\rangle}{\|{\bf W}_0\| \|{\bf W}_j\|} = \frac{4}{3N} \sum_j \langle {\bf W}_0 \cdot {\bf W}_j\rangle
	\label{eq:PseudospinCorrelations_apx}
\end{equation}

\begin{figure}[h]
    \centering
    \includegraphics[ width=0.5\columnwidth]{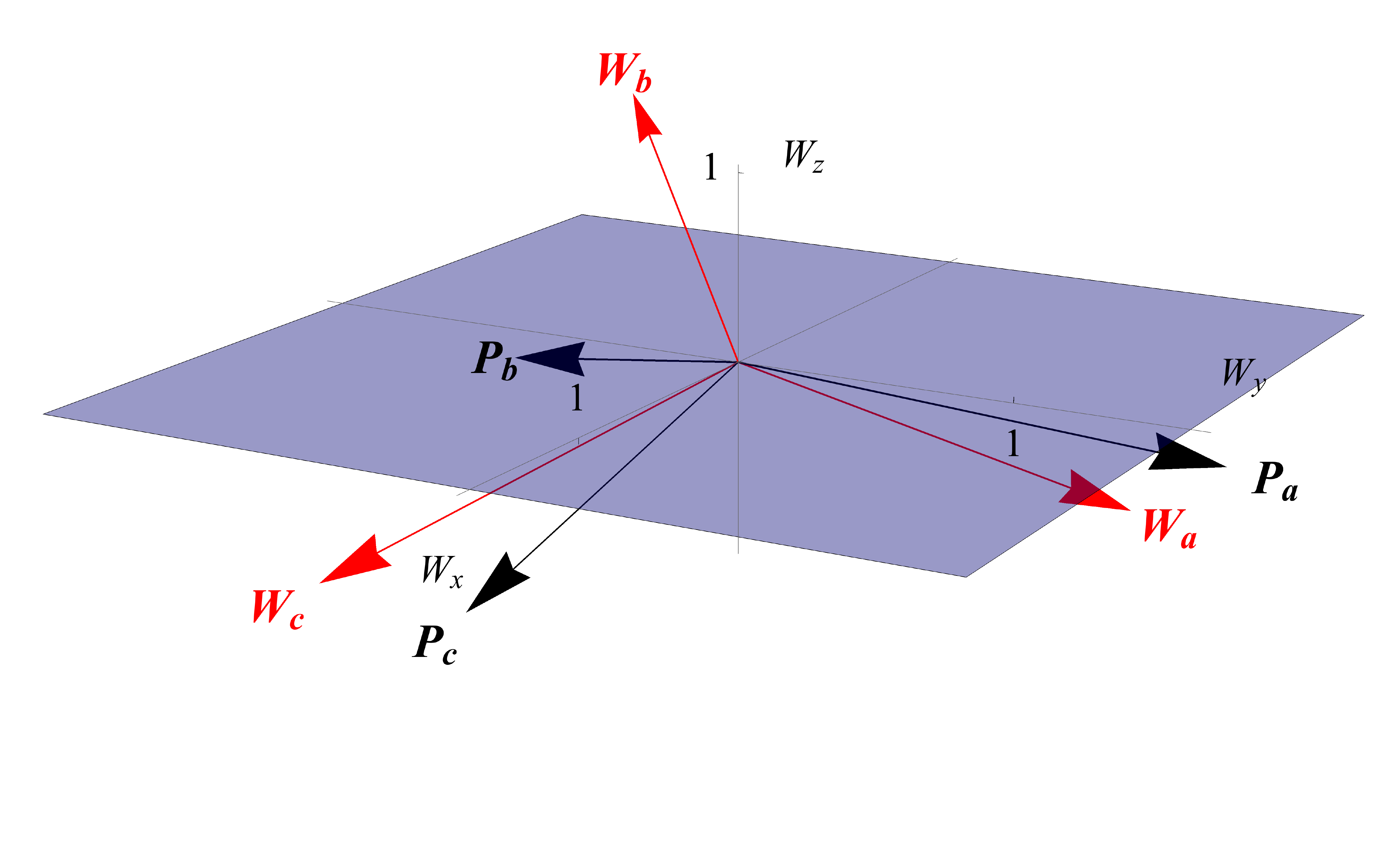}  
      \caption{{\bf Pseudospin vectors.}  The triplet of pseudospin vectors that preserves the local-divergence free condition (red) is transformed to a set of vectors that are rotationally symmetric around the $W_z$-axis (black). Depending on the (an)isotropy of the bond couplings, different subsets are selected at low temperatures: (i) all three (isotropic point), (ii) ${\bf P}_b$ and ${\bf P}_c$ ($J_z < 1/3$), (iii) only ${\bf P}_a$  ($J_z > 1/3$).
		  }
    \label{fig:Supp_PseudoSpinVectors}
\end{figure}

\section{Sign-free quantum Monte Carlo simulations}

Our large-scale quantum Monte Carlo simulations are performed in the parton basis of the spin system \cite{Kitaev2006anyons}, in which they are sign-problem free \cite{Nasu2014vaporization}. Technically, a single Monte Carlo step consists of a local update of the $\mathbb{Z}_2$ gauge degree of freedom on a random lattice bond, followed by an evaluation of the Majorana (free) energy required for the calculation of a Metropolis acceptance probability. While this latter step can be accomplished using exact diagonalization (ED) techniques \cite{Nasu2014vaporization}, a far more efficient algorithm that avoids the cubic scaling of the ED approach is to employ
the Green-Function-Based Kernel Polynomial Method (GF-KPM) \cite{Weisse2009}, which was recently introduced for Kitaev systems \cite{Mishchenko2017}. Using this apporoach, the free energy {\em change} of the flip of a $\mathbb{Z}_2$ gauge variable is calculated from a set of four Green functions which can be approximated with a limited number of Chebyshev polynomials (typically with 128 to 256 coefficients in our case). The computational cost of each MC update is thereby reduced to $\mathcal{O}(N)$ scaling -- a crucial step which enables us to simulate relatively large system sizes with up to 1728 spins.

Note that for the transformation of the spin model to the Majorana basis, we follow the local approach originally pioneered by Kitaev \cite{Kitaev2006anyons}. In this local approach all lattice bonds carry gauge field degrees of freedom and can be addressed by a single-flip MC step, in contrast to an alternative scheme which makes use of a Jordan-Wigner transformation and where the resulting $\mathbb{Z}_2$ gauge field degrees of freedom are located only on the $z$-bonds \cite{Nasu2014vaporization}. A crucial advantage of the local approach is that it allows us to simulate systems with periodic boundary conditions in {\em all} directions, thereby minimizing finite-size effects. There remains one subtlety, however:  By replacing each spin with four Majorana fermions, the Kitaev ansatz artificially expands the Hilbert space of the system, which would normally require a rigorous projection into the physical subspace. However, it is known that physical and unphysical states for a given lattice and $\mathbb{Z}_2$ gauge field configuration are distinguished by their fermionic parity, reducing the effect of states of the expanded Hilbert space to deviations of order $1/N$, which can be neglected in the thermodynamic limit \cite{Zschocke2015}. In practical terms, we have checked that already for small system sizes of  $N=64$ sites, we cannot establish differences between the two approaches up to our numerical efficiency.

Finally, in order to prevent our MC simulation from  freezing at low temperatures, we have applied a parallel tempering scheme \cite{Hukushima1996} with one replica exchange proposed between every pair of neighboring temperature points after every MC sweep. This enabled us to perform simulations down to temperatures of  $\mathcal{O}(10^{-3})$ in terms of the magnetic coupling strength. To calculate observables on the lower end of this limit and to interpolate between temperature points, we have used the Ferrenberg-Swendsen reweighting method with multiple histograms \cite{FerrenbergSwendsen88,FerrenbergSwendsen89}.